\documentstyle[mnras_cite,epsfig,eqnarray]{mn2e}
\title[Beam Power and Radio Power]{The Relationship Between Beam Power and Radio Power for Classical Double Radio Sources}
\author[R. A. Daly et al.]
{Ruth.~ A.~ Daly$^{1}$,\thanks{E-mail:
rdaly@psu.edu}
Trevor~B.~ Sprinkle$^{1}$, 
Christopher~ P.~ O'Dea$^{2,3}$, Preeti~ Kharb$^{2}$, 
\newauthor and 
Stefi A. Baum$^{4,5}$ \\
$^{1}$Penn State University, Berks Campus, Reading, PA 19608, USA \\
$^{2}$Department of Physics,
Rochester Institute of Technology, 54 Lomb Memorial Drive,
Rochester, NY 14623, USA\\
$^{3}$Harvard Smithsonian Center for Astrophysics, 60 Garden St.
Cambridge, MA 02138\\
$^{4}$Center for Imaging Science,
Rochester Institute of Technology, 54 Lomb Memorial Drive,
Rochester, NY 14623, USA\\
$^{5}$Radcliffe Institute for Advanced Study, 10 Garden St. Cambridge,
MA 02138}

\begin{document}



\maketitle

\label{firstpage}
 
\begin{abstract}
Beam power is a fundamental parameter that describes, in part,  
the state of a supermassive black hole system. Determining the
beam powers of powerful classical double radio sources requires
substantial observing time, so it would be useful to determine
the relationship between beam power and radio power so that 
radio power could be used as a proxy for beam power. 
A sample of 31 powerful classical double radio sources with 
previously determined beam and radio powers 
are studied; the sources have redshifts between about 0.056 and 1.8. 
It is found that the relationship between 
beam power, $L_j$, and radio power, $P$, is well described by 
$\log{L_j} \approx 0.84 (\pm 0.14) \log{P} + 2.15 (\pm 0.07)$, 
where both $L_j$ and $P$ are in units of $10^{44} \rm{~erg~ s}^{-1}$. 
This indicates that beam power is converted to radio power with 
an efficiency of about 0.7\%. The ratio of beam power to radio power is  
studied as a function of redshift; there is no significant evidence for 
redshift evolution of this ratio over the redshift range studied. 
The relationship is consistent  with 
empirical results obtained by Cavagnolo et al. (2010) for radio 
sources in gas rich environments, which are primarily FRI sources, and
with the theoretical predictions of Willott et al. (1999).

\end{abstract}


\section{INTRODUCTION}
A powerful classical double radio source, also known as an FRII source
(Fanaroff \& Riley 1974), is powered by large scale outflows from  
a supermassive black hole system that resides at the center of a
galaxy (e.g. Blandford \& Rees 1974; Scheuer 1974).  
The energy per unit time, or beam power, carried from the vicinity
of the supermassive black hole to the large scale radio source is a 
fundamental physical parameter that, in part, describes the physical
state of the black hole system. The beam power can be used to study 
many aspects of the source, 
the source population, interactions of the source with its 
environment, and the role of feedback between the black hole system, its
environment, and the host galaxy
(e.g. Silk \& Rees 1998; 
Willott et al. 1999; Eilek \& Owen 2002; Birzan et al. 2004; 
Croston et al. 2005; Dunn et al. 2005; Dunn \& Fabian 2006; 
Rafferty et al. 2006;  Belsole et al. 2007; Jetha et al. 2007; 
Birzan et al. 2008; Cavagnolo et al. 2008; 
Daly 2009, 2011; Cavagnolo et al. 2010; 
Mart\'{i}nez-Sansigre \& Rawlings 2011).  

It is possible to determine the beam power of a classical double (FRII) radio
source using multifrequency radio data that covers most of the radio emitting
region and resolves several different regions of each radio lobe 
(e.g. Rawlings \& Saunders 1991; Daly 1994; 
Kharb et al. 2008; O'Dea et al. 2009; Daly et al. 2010); see
O'Dea et al. (2009) for a detailed discussion of the method. 
Obtaining these data requires a considerable amount of observing time. 
If a relationship between beam power and radio power can be 
established, it can be applied to other FRII sources so that their beam
power can be estimated from their radio power.

The relationship between beam power and radio power has been studied for 
samples of radio sources that reside in gas rich environments. 
These sources are primarily FRI sources (Fanaroff \& Riley 1974) 
at relatively low redshift (Cavagnolo et al. 2010; 
Birzan et al. 2008; Rafferty et al. 2006; Dunn et al. 2005; 
Birzan et al. 2004). It has also been studied for a small sample of 14
FRII galaxies in the context of an open cosmological model (Wan \& Daly 1998). 

Here, the relationship between beam power and radio power is determined
for a sample of 31 FRII radio galaxies with a broad range of redshift. 
The sample is described in section 2. The results are presented in 
section 3, discussed and compared with other results and model predictions 
in section 4, and summarized in section 5. A standard spatially flat 
cosmological model 
with normalized mean mass density $\Omega_m = 0.3$, cosmological constant
$\Omega_{\Lambda} = 0.7$, and Hubble constant $\rm{H_0} = 70~ \rm { km~s}^{-1} 
\rm{ Mpc}^{-1}$ is assumed throughout. 

\section{SAMPLE}
The sample studied consists of 31 very powerful FRII radio galaxies with 
redshifts ranging from about 0.056 to 1.79 and core-hot spot sizes
ranging from about 30 to 400 kpc drawn from the sample studied by 
O'Dea et al. (2009). The sources have both beam powers
and radio powers determined, and are 3CRR sources  
(Laing, Riley, \& Longair 1983). 
The sources and source properties 
are listed in Table 1. The intrinsic 178 MHz 
energy per unit time per unit frequency, 
$P_{\nu}(178~\rm{MHz})$,   
is listed in column 3, and is determined from the 178 MHz flux density, 
$f_{\nu}$, and the radio spectral index, $\alpha$, defined as $f_{\nu} \propto
\nu^{-\alpha}$.  Radio flux densities 
and spectral indices are obtained from Jackson \& 
Rawlings (1997) and Willott et al. (1999).
The 178 MHz radio power, $P_{178}$, is obtained 
by multiplying $P_{\nu}(178~\rm{MHz})$ by 178 MHz, and the integrated radio 
power, $P_{int}$ is obtained by integrating $P_{\nu}$ over
the frequency range from 200 to 400 MHz using the radio spectral index listed
in column 9.  This frequency range is selected to match that studied by 
Cavagnolo et al. (2010) so that the results can be easily compared.
Values of 
$P_{178}$ and $P_{int}$ are listed in columns 4 and 5, respectively, 
of Table 1. Beam
powers, $L_j$, are the weighted sum of the beam power from 
each side of a given source 
obtained from O'Dea et al. (2009) and are listed in column 6. 
The ratio of the beam power to the radio power is listed in columns 
7 and 8 for $P_{178}$ and $P_{int}$, respectively. 

The classical double radio galaxies studied here 
are among the most powerful extended radio sources 
known in the cosmos. These powerful radio sources
are often much larger than their parent galaxies and 
have cigar-like radio lobe structure
indicating that the forward region of the radio emitting lobe is 
moving into the ambient gas supersonically (e.g. Alexander \& Leahy 1987;
Leahy, Muxlow, \& Stephens 1989; Liu, Pooley, \& Riley 1992). Since
the structure of the sources indicates that each is 
growing at a supersonic rate, the equations of
strong shock physics may be applied to the sources 
(Leahy, Muxlow, \& Stephens 1989).
As discussed in detail by Leahy (1990), 
Rawlings \& Saunders (1991), Daly (1990, 1994), 
Wan \& Daly (1998), 
Wan et al. (2000), and O'Dea et al. (2009), 
for example, this means that the beam power 
$L_j = \kappa_L v P^{\prime} a^2$, where 
$\kappa_L$ is a constant, $v$ is the rate of growth of
the source, $a$ is the radius of the cross-sectional area
of the forward region of the shock, and $P^{\prime}$ is the postshock 
pressure. Each of these parameters has been determined empirically.  
O'Dea et al. (2009) provide a detailed
description of the method and related uncertainties and 
a comprehensive list of beam powers obtained using this method; 
these are the sources studied here. Remarkably, 
O'Dea et al. (2009) found that the beam power is independent
of offsets from minimum energy conditions due to a cancellation
of the way the offset enters into the determinations of $P^{\prime}$ and $v$, 
so the beam power is insensitive to
assumptions regarding minimum energy conditions. 
  
The beam power studied here has been obtained using the equation
$L_j = \kappa_L v P^{\prime} a^2$, where $\kappa_L = 4 \pi$ (e.g. Leahy 1990; 
Wan et al. 2000). The rate of growth of the source $v$ is obtained 
using the equation $v = \Delta x/\Delta t$.   The change in the 
radio spectrum across a region $\Delta x$ along the symmetry axis of the 
source is used to 
obtain the time  $\Delta t$ that has elapsed for the source length to 
change by an amount $\Delta x$. 
The spectral aging method of Meyers \& Spangler (1985) was
applied to the Kardashev-Pacholczyk and Jaffe-Perola models 
(Jaffe \& Perola 1973) to obtain $\Delta t$, 
and similar results were obtained with both models. 
Details of the method used to 
obtain the quantities described above and  
the computations of their uncertainties are given in
sections 2.1, 2.2, 3.1, 3.2, and 3.3 of O'Dea et al. (2009). 
In brief, 
uncertainties in the 
injection spectral index, and the flux density at each frequency studied
are included in the uncertainty of $\Delta t$ obtained for 
each value of $\Delta x$. Several different values
of $\Delta x$ were considered for each source, and the value of 
$\Delta x/\Delta t$ obtained was found to be 
independent of the value of $\Delta x$. 
In the computation of the 
rate of growth of the source, offsets of the true magnetic field
strength $B$ 
from minimum energy conditions 
were parameterized by $B = b B_{min}$ where $B_{min}$ is the 
minimum energy magnetic field, and the geometric mean of the 
minimum energy fields 10 and 25 kpc from the hot spot along the 
symmetry axis of the source were used
to obtain $B_{min}$ over the spectral aging region. 
The postshock pressure $P^{\prime}$ was taken 
to be the pressure in the region 10 kpc behind the hot spot moving toward
the core of the host galaxy along the symmetry axis of the radio source; 
the pressure 
can be written $P^{\prime} = (1.33b^{-1.5}+b^2)B^2_{min}/(24\pi)$ (e.g. Wan et
al. 2000), where $B_{min}$ here is obtained 10 kpc from the hot spot. 
It is shown in section 3.3 of O'Dea et al. (2009) that the 
offset from minimum energy conditions entering through $v$ cancels that
entering through $P^{\prime}$ so that, for the sources studied here, 
$L_j$ is independent of offsets from minimum energy conditions. 
The uncertainty of the beam power is dominated by the 
uncertainties of the source rate of growth and pressure.
The half-width of the source $a$ is taken to be the  
deconvolved half-width of the source, which is equal to 
$3^{-1/2}$ times the deconvolved FWHM of the source 
(see, for example, section 5.1 of Wellman et al. 1997). 
Widths were extracted at 1.4 GHz a distance 10 kpc behind the 
hot spot.  Detailed studies of these sources show that the source
width is independent of the frequency of observation 
for the range of frequencies studied (Daly et al. 2010). 
For the new sources studied by O'Dea et al. (2009), the 
widths were obtained using a cross-sectional slice taken perpendicular
to the symmetry axis of the source, as described by Daly et al. (2010).
The deconvolved FWHM, $w_t$, is given by $w_t = (w_G^2 -w_b^2)$, where 
$w_G$ is the FWHM of the best fit Gaussian to the slice and $w_b$ is 
the observing beamwidth. The widths obtained were found to be in 
good agreement with those obtained earlier by Wellman et al. (1997). 
The uncertainty of the deconvolved FWHM, 
$\delta w_t$, was obtained using the expression  
$(\delta w_t)^2 = (w_G/w_t)^2(\delta w_G)^2 
+(w_b/w_t)^2(\delta w_b)^2$, where $\delta w_G$ is the uncertainty of
the FWHM of the best fit Gaussian profile and $\delta w_b$ is 
the uncertainty of the observing beam.  In practice, $\delta w_t$
was always very close to $\delta w_G$; only widths very close to the
hot spot had a larger uncertainty, and these were at most 
about 35 \% larger than $\delta w_G$. 

\section{RESULTS}

The beam power was studied as a function of radio power for the full 
sample of 31 sources and for a sample of 30 galaxies obtained by 
excluding the lone low redshift source in the sample,
Cygnus A. Very similar results are obtained using the radio power 
defined by $P_{178}$ and $P_{int}$, and considering the samples of 
30 and 31 sources. The results are illustrated in Figure 1, 
and best fit parameters are listed in 
Table 2.  The uncertainties on the best fit parameters have been 
adjusted to bring the reduced chi-square 
of the fit to unity.  Overall, the data are described by the equation
\begin{equation}
\log(L_{44}) = 0.84 (\pm 0.14) 
\log(P_{44})+ 2.15 (\pm 0.07)~,
\end{equation}
where $L_{44} \equiv L_j/(10^{44} \rm{erg~s}^{-1})$, and 
$P_{44} \equiv P/(10^{44} \rm{erg~s}^{-1})$. 
For powerful FRII sources, most of the radio power is produced by the 
outer half of the radio lobe (near the radio hot spots) (e.g. Leahy, Muxlow,
\& Stephens 1989).  The result obtained
here, $L_{44} \approx (140 \pm 20) P_{44}^{0.84 \pm 0.14}$,
is consistent with $L_j \propto P$ at about one sigma. 
These results are consistent with those
obtained with a sample of 14 radio galaxies (Wan \& Daly 1998).
The efficiency with which beam power is converted to radio power
is $\epsilon = P/L_j \approx 0.007 P_{44}^{0.16 \pm 0.14}$, and 
the constant of proportionality 
suggests the beam power may be converted to radio power with an overall 
efficiency of about 0.7 \%.

The radio power and the beam power each increase with redshift, so 
there is a concern that the intrinsic relationship between beam power and
radio power may be masked because each is correlated with redshift. 
This effect can be circumvented by studying the ratio 
$L_j/P$ rather than studying each quantity separately. 
To explore this, the 
ratio of the beam power to the radio power, $L_j/P$, is studied
as a function of redshift with and without Cygnus A. 
The results are illustrated in Figure 2, and best fit parameters are
listed in Table 2.   
Without Cygnus A, the ratio $L_j/P$ is independent of redshift,
indicating that $L_j \propto P$ is intrinsic to the sources. 
When Cygnus A is included, there is a tendency for the ratio to
increase with redshift though the significance of the 
result is only about one sigma. This suggests that 
the relationship between beam power
and radio power obtained for powerful classical double radio sources 
is intrinsic to the sources and 
is independent of redshift for redshifts ranging from 
about 0.4 to 2, and possibly for redshifts from about zero to 2. 
 
The expression used to obtain the beam power can be compared with that
used to obtain the radio power of a given source. The radio power 
is obtained by integrating the radio surface brightness of the 
the source to obtain the flux density, multiplying this by the 
luminosity distance squared and appropriate redshift factors, 
and multiplying by the rest frame
radio frequency or integrating over radio frequency to obtain
$P_{178}$ or $P_{int}$, respectively. The beam power depends on 
$L_j \propto \Delta x \nu_T^{1/2} a_{10}^{17/14} a_{25}^{-3/14} 
S_{10}^{11/14} S_{25}^{3/14} (1+z)^{(3+ \alpha)}$, where 
$a_{10}$ and $a_{25}$ are the widths and 
$S_{10}$ and $S_{25}$ are the mean surface brightnesses of 
cross-sectional slices of the radio lobe 10 and 25 kpc, respectively,
from the hot spot along the symmetry axis of the source and
$\nu_T$ is the spectral aging break frequency. 
This expression assumes that synchrotron cooling dominates over 
inverse Compton cooling and that $b \leq 1$, both 
of which apply to the sources studied here (O'Dea et al. 2009). 
As noted above,
the length 
$\Delta x$ is measured from the hot spot to the location along the 
symmetry axis of the lobe where the break frequency $\nu_T$ is measured, 
and the product 
$\Delta x \nu_T^{1/2}$ was found to be independent of 
$\Delta x$, thus, the beam power is also independent of $\Delta x$ 
(O'Dea et al. 2009). 
The beam power depends upon several factors that do not enter
into the calculation of the radio power, and the radio power depends
upon factors that do not enter the beam power, so the correlation
obtained here between these two parameters is likely to characterize an
intrinsic physical relationship between beam power and radio power.  
In addition, the results obtained here are consistent with those 
obtained by Cavagnolo et al. (2010) who used a
completely different and independent method of determining the beam power, 
as discussed in section 4.

\begin{figure}
    \centering
    \includegraphics[width=80mm]{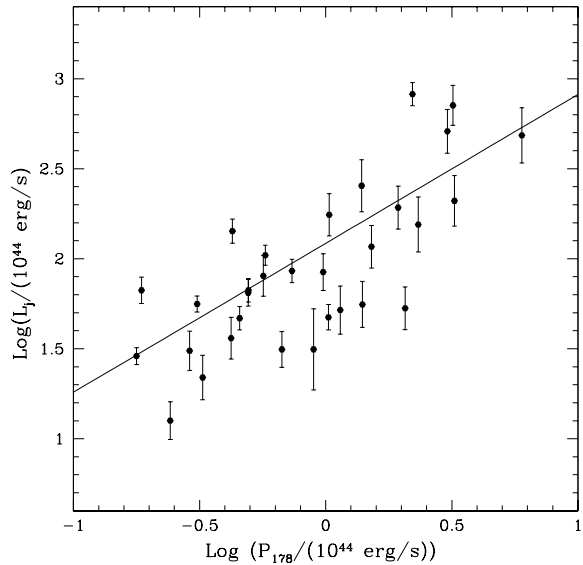}
			 
\caption{Beam power as a function of radio power for the 31 radio galaxies 
in the sample. 
The radio power shown is obtained at 178 MHz as described in the text. 
The best fit line to this data set is shown, and  
parameter values are listed in Table 2. }
		  \label{fig:F1}
    \end{figure}

\begin{figure}
    \centering
    \includegraphics[width=80mm]{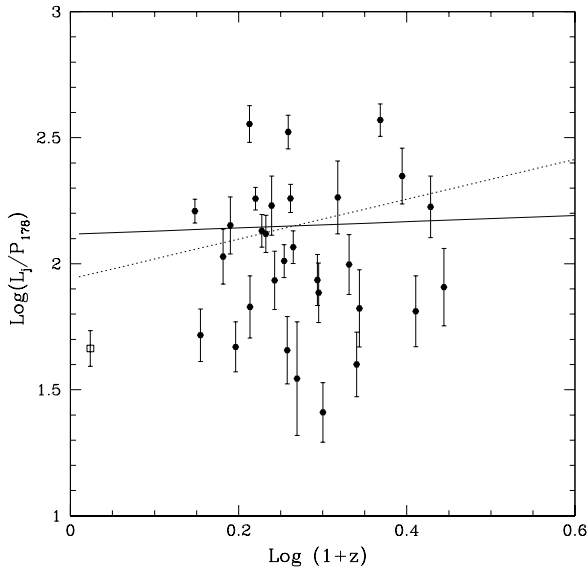}
\caption{The ratio of the beam power to the radio power obtained at 
178 MHz as a function of redshift. Thirty of the radio galaxies are 
indicated as filled circles; Cygnus A is shown as an open square.
The best fit line to 30 radio galaxies is indicated as a solid line and 
that for all 31 sources is shown as a dotted line. The best fit
parameters are listed in Table 2.  }
		  \label{fig:F2}
    \end{figure}

\section{DISCUSSION}
The results obtained here can be compared with those obtained by 
Cavagnolo et al. (2010) who studied radio 
sources in gas rich clusters of galaxies and other gas rich 
environments. Cavagnolo et al. (2010)
expanded upon the sample studied by Birzan et al. (2008) by adding 
very lower power radio sources to the sample. The sources studied
by these groups are primarily FRI sources.
The method used to obtain the beam power for these sources 
is significantly different from that described in section 3. For the 
radio sources in gas rich environments 
studied by Birzan et al. (2008) and Cavagnolo et al. (2010), 
the P dV work done by 
the relativistic plasma on the intracluster medium is obtained 
and combined 
with the buoyancy timescale of the radio source to obtain the beam power
(e.g. Fabian et al. 2006; Rafferty et al. 2006). 
Cavagnolo et al. (2010) obtain $L_{44} \approx 0.58 (10^4~P_{44})^{0.7}$, 
or $L_{44} \approx 366 P_{44}^{0.7}$, and  
the coefficient may be approximated as $370 \pm 130$.
As noted by these authors, the buoyancy timescale is likely to be an 
overestimate of the true source age and some of the sources may be 
affected by multiple outflows; these uncertainties 
would increase the uncertainties on parameters stated above.  

The slope of the relationship obtained here, $0.84 \pm 0.14$, is 
within one sigma of that 
obtained by Cavagnolo et al. (2010), $0.70 \pm 0.12$. 
The normalization of the relationship 
obtained by Cavagnolo et al. (2010),  $370 \pm 130$, 
differs from that obtained here, $140 \pm 20$ by
less than 2 sigma. Thus, the results
obtained are remarkably similar even though 
different types of radio sources at different redshifts are studied. 
The small differences that do exist   
may be accounted for by uncertainties in fitted 
parameters, or by intrinsic differences between the sources. More data will  
be needed to compare results obtained with 
different methods in more detail. 

The agreement between the results obtained here and those of 
Cavagnolo et al. (2010) is particularly remarkable because the interaction
of the sources with their environments is quite different: the forward 
region of FRII sources is moving supersonically with respect to 
the ambient gas resulting in a strong shock whereas FRI sources
are not moving supersonically and in any case the method used
by Cavagnolo et al. (2010) does not include energy that has gone into shocks.
The results of 
Cavagnolo et al. (2010) described above 
imply an efficiency of conversion of beam power to radio
power of $\epsilon = P/L_j \approx 0.003P_{44}^{0.3 \pm 0.12}$; the slope
is consistent with zero at about 2.5 sigma and the coefficient suggests
a conversion efficiency of about 0.3\%. If the slope is taken to be
non-zero, it would suggest that the conversion efficiency increases
with radio power and is about 0.3\% at a power of $10^{44} \rm{erg/s}$. 
Thus, both types of source appear to convert
beam power to radio power with similar efficiencies: 
about 0.7\% for FRII
sources and about 0.3\% for FRI sources. This provides a hint that
the physics of the conversion of
beam power to relativistic electrons and magnetic fields may be
similar for FRI and FRII sources, perhaps with the efficiency of
conversion increasing with radio power for FRI sources. 
Alternatively, it is possible that 
different physical
processes for the conversion of beam power to radio power have 
similar efficiencies. 

The results obtained here can be 
compared with the theoretical model described by 
Willott et al. (1999), who found that 
$L_j \approx 1.7 \times 10^{45}f^{3/2} 
(P_{44})^{6/7} \rm{erg ~s}^{-1}$, where the 
term $f$ is expected to lie between 1 and 20 and is a combination
of factors including the filling factor of the lobes, departure from 
minimum energy conditions, and the fraction of energy in non-radiating
particles. This model prediction is obtained 
in the context of a detailed model based on the work of Falle (1991) and
Kaiser \& Alexander (1997) for extended radio sources.
The dependence of beam power on radio power predicted by the model, 
which has an exponent of $0.86$,  
is remarkably close to the value of $0.84$ obtained here. 
The normalization of the relationship obtained here,  
$L_{44} \approx 140 P_{44}^{0.84}$, indicates a value of
$f \approx 4$. The fact that both the results 
obtained here and
those obtained by Cavagnolo et al. (2010) are in fairly good agreement
with the model predictions suggests that the fundamental model provides a 
good description of the sources. 

\section{SUMMARY}
A sample of 31 powerful FRII radio galaxies is studied to determine
the relationship between beam power and radio power for sources of this
type. The data indicated that $\log(L_{44}) = 0.84 (\pm 0.14) 
\log(P_{44})+ 2.15 (\pm 0.07)$,
where $L_{44} \equiv L_j/(10^{44} \rm{erg~s}^{-1})$, and 
$P_{44} \equiv P/(10^{44} \rm{erg~s}^{-1})$.
The results listed in Table 2 may be applied to samples of FRII sources
with known radio power to obtain estimates of the beam power. 

To determine whether the relationship between
beam power and radio power evolves with redshift, the ratio of
the beam power to radio power was studied as a function of redshift 
with and without the one low redshift source, Cygnus A. Excluding 
Cygnus A, there is no redshift evolution of the ratio of beam power to radio 
power. Including Cygnus A, there is a tendency for the ratio to increase
with redshift, but the significance of the result is marginal.  
Thus, there is no
conclusive evidence at this time that the ratio evolves with redshift. 

The results obtained here are broadly consistent with those obtained
by Cavagnolo et al. (2010) for samples of radio sources in 
gas rich environments, which consist primarily of 
low redshift FRI sources.   
Interestingly, the results obtained 
here are consistent with the radio power being proportional to the 
beam power, with an efficiency of conversion of beam power to radio 
power of about 0.7\% that is independent of or very weakly dependent
upon radio power. This is comparable to the efficiency of about 0.3\% 
indicated by FRI sources, which may be weakly dependent upon
radio power in the sense that the efficiency of conversion of beam power
to radio power may increase weakly with radio power. 

The relationship between beam power and radio power obtained 
here may be compared with theoretical predictions obtained in 
the context of a specific model 
for the sources. The agreement between the results obtained
here and the detailed model described by Willott et al. 
(1999) is quite impressive.

\section*{Acknowledgments}We would like to thank the referee
for very helpful comments and suggestions. This work is supported in part by 
Penn State University (R.A.D. and T.B.S.) 
and the Radcliffe Institute for Advanced
Study at Harvard University (S.A.B.).


\begin{table*}
\begin{minipage}{140mm}
\caption{Radio and Beam Powers of FRII Sources}   
\label{tab:comp}        
\begin{tabular}{llccccccc}   
\hline\hline                    

Source  &z&$P_{\nu}(178~ \rm{MHz})$&$P_{178}$&$P_{\rm{int}}$&
$L_j$
\footnote{Total beam powers are the weighted sum of the 
beam power from each of the two lobes of each source. Input values for 
each side of each radio galaxy (RG) are obtained from by O'Dea et al. (2009), 
who used the data sets 
of Leahy, Muxlow, \& Stephens (1989), Liu, Pooley, \& Riley (1992), 
Guerra, Daly, \& Wan (2000), and new observations. The beam power is
obtained by applying the equations of strong shock physics to the
source. }   
&${L_j} / {P_{178}}$&${L_j} / {P_{\rm{int}}}$

& $ {\alpha}$ \\
&&$10^{36}~ \rm{erg~ s}^{-1} \rm{Hz}^{-1}$&$10^{44}~ \rm{erg~ s}^{-1}$ &
$10^{44}~ \rm{erg~ s}^{-1}$ & $10^{44}~ \rm{erg~ s}^{-1}$ \\
(1)    &    (2)         &    (3)          &   (4)           &  (5) 
      &       (6)    &(7)&(8)&(9)         \\
\hline                          
			
3C	6.1	&	0.840	&	0.413	&	0.735	&	0.592	&$	86	\pm	13	$&$	116	\pm	17	$&$	144	\pm
22	$&	0.68	\\							
3C	34	&	0.689	&	0.278	&	0.494	&	0.333	&$	67	\pm	10	$&$	135	\pm	20	$&$	200	\pm
30	$&	1.06	\\							
3C	41	&	0.795	&	0.256	&	0.456	&	0.398	&$	47	\pm	7	$&$	102	\pm	15	$&$	117	\pm	18
$&	0.51	\\							
3C 	44	&	0.660	&	0.174	&	0.309	&	0.212	&$	56	\pm	6	$&$	181	\pm	19	$&$	264	\pm
27	$&	1.02	\\							
3C	54	&	0.827	&	0.324	&	0.576	&	0.404	&$	105	\pm	13	$&$	182	\pm	23	$&$	259	\pm
33	$&	0.98	\\							
3C	55	&	0.735	&	0.580	&	1.03	&	0.702	&$	176	\pm	48	$&$	170	\pm	46	$&$	250	\pm
68	$&	1.04	\\							
3C	68.2	&	1.575	&	1.82	&	3.24	&	2.19	&$	210	\pm	68	$&$	65	\pm	21	$&$	96	\pm	31
$&	1.05	\\							
3C 	114	&	0.815	&	0.240	&	0.427	&	0.280	&$	142	\pm	22	$&$	333	\pm	51	$&$	508
\pm	78	$&	1.12	\\							
3C	142.1	&	0.406	&	0.100	&	0.178	&	0.134	&$	29	\pm	3	$&$	162	\pm	18	$&$	214	\pm
23	$&	0.82	\\							
3C	169.1	&	0.633	&	0.105	&	0.186	&	0.136	&$	67	\pm	11	$&$	358	\pm	60	$&$	489
\pm	82	$&	0.88	\\							
3C	172	&	0.519	&	0.162	&	0.289	&	0.213	&$	31	\pm	8	$&$	107	\pm	27	$&$	144	\pm
36	$&	0.86	\\							
3C	239	&	1.781	&	3.37	&	5.99	&	4.00	&$	484	\pm	170	$&$	81	\pm	29	$&$	121	\pm	43
$&	1.08	\\							
3C	244.1	&	0.428	&	0.136	&	0.242	&	0.183	&$	13	\pm	3	$&$	52	\pm	13	$&$	69	\pm
17	$&	0.82	\\							
3C	247	&	0.749	&	0.237	&	0.422	&	0.351	&$	36	\pm	10	$&$	86	\pm	23	$&$	103	\pm
27	$&	0.61	\\							
3C	265	&	0.811	&	0.642	&	1.14	&	0.807	&$	52	\pm	16	$&$	45	\pm	14	$&$	64	\pm	20
$&	0.96	\\							
3C	267	&	1.144	&	1.09	&	1.94	&	1.39	&$	192	\pm	53	$&$	99	\pm	27	$&$	139	\pm	38
$&	0.93	\\							
3C	268.1	&	0.973	&	0.853	&	1.52	&	1.28	&$	117	\pm	32	$&$	77	\pm	21	$&$	91	\pm
25	$&	0.59	\\							
3C	280	&	0.996	&	1.16	&	2.06	&	1.56	&$	53	\pm	14	$&$	26	\pm	7	$&$	34	\pm	9	$&
0.81	\\							
3C	289	&	0.967	&	0.550	&	0.978	&	0.741	&$	84	\pm	20	$&$	86	\pm	20	$&$	114	\pm
27	$&	0.81	\\							
3C	322	&	1.681	&	1.70	&	3.03	&	2.30	&$	510	\pm	140	$&$	168	\pm	47	$&$	222	\pm
62	$&	0.81	\\							
3C	324	&	1.206	&	1.31	&	2.33	&	1.69	&$	155	\pm	55	$&$	67	\pm	23	$&$	92	\pm	32
$&	0.9	\\							
3C	325	&	0.860	&	0.504	&	0.896	&	0.715	&$	31	\pm	16	$&$	35	\pm	18	$&$	44	\pm	23
$&	0.7	\\							
3C	330	&	0.549	&	0.318	&	0.566	&	0.450	&$	80	\pm	21	$&$	142	\pm	37	$&$	179	\pm
47	$&	0.71	\\							
3C	337	&	0.635	&	0.182	&	0.325	&	0.268	&$	22	\pm	6	$&$	67	\pm	19	$&$	82	\pm	23
$&	0.63	\\							
3C	356	&	1.079	&	0.780	&	1.39	&	0.953	&$	254	\pm	85	$&$	183	\pm	61	$&$	267	\pm
89	$&	1.02	\\							
3C	427.1	&	0.572	&	0.377	&	0.670	&	0.471	&$	31	\pm	7	$&$	47	\pm	11	$&$	67	\pm
15	$&	0.97	\\							
3C	437	&	1.480	&	1.79	&	3.19	&	2.44	&$	711	\pm	180	$&$	223	\pm	57	$&$	291	\pm
74	$&	0.79	\\							
3C	441	&	0.708	&	0.277	&	0.493	&	0.370	&$	65	\pm	11	$&$	131	\pm	22	$&$	175	\pm
30	$&	0.83	\\							
3C	469.1	&	1.336	&	1.24	&	2.21	&	1.56	&$	820	\pm	120	$&$	371	\pm	55	$&$	526	\pm
78	$&	0.96	\\							
3C	194	&	1.190	&	0.786	&	1.40	&	1.06	&$	56	\pm	16	$&$	40	\pm	12	$&$	52	\pm	15
$&	0.8	\\							
3C	405	&	0.056	&	0.576	&	1.03	&	0.781	&$	47	\pm	8	$&$	46	\pm	7	$&$	61	\pm	10	$&
0.8	\\

\end{tabular}
\end{minipage}
\end{table*}

\begin{table*}
\begin{minipage}{140mm}
\caption{Best Fit Parameters}   
\label{tab:comp}        
\begin{tabular}{ccccc}   
\hline\hline                    
N&A
\footnote{Fits are to the equation $\log(A) = m~\log(B)+b$. 
The uncertainties of the best fit parameters have been adjusted to 
account for deviations of the reduced $\chi^2$ from unity.}
&B&m&b\\
\hline 
31&$L_j/(10^{44}~ \rm{erg~s}^{-1})$&$ P_{178}/(10^{44}~ \rm{erg~s}^{-1})$&$0.83 \pm 0.14$&$
2.09 \pm 0.06$\\

30\footnote{The one low redshift source, Cygnus A (3C 405) has been excluded 
from the analysis bringing the number of sources to 30.}
&$L_j/(10^{44}~ \rm{erg~s}^{-1})$&$ P_{178}/(10^{44}~ \rm{erg~s}^{-1})$&$0.87 \pm 0.13$&$
2.12 \pm 0.06$\\

31&$L_j/(10^{44}~ \rm{erg~s}^{-1})$&$ P_{\rm{int}}/(10^{44}~ \rm{erg~s}^{-1})$&$0.80 \pm 0.14$&$ 2.18 \pm 0.07$\\

30&$L_j/(10^{44}~ \rm{erg~s}^{-1})$&$ P_{\rm{int}}/(10^{44}~ \rm{erg~s}^{-1})$&$0.84 \pm 0.14$&$
2.22 \pm 0.07$\\

31&$L_j/P_{178}$&$(1+z)$&$0.79 \pm 0.60$&$1.94 \pm 0.15$\\
30&$L_j/P_{178}$&$(1+z)$&$0.12 \pm 0.71$&$2.12 \pm 0.19$\\
31&$L_j/P_{\rm{int}}$&$(1+z)$&$0.84 \pm 0.63$&$2.06 \pm 0.16$\\	
30&$L_j/P_{\rm{int}}$&$(1+z)$&$0.16 \pm 0.75$&$2.24 \pm 0.20$\\				
\end{tabular}
\end{minipage}
\end{table*} 

\end{document}